      [1999/12/01 v1.4c Il Nuovo Cimento]
\documentclass{cimento}

\usepackage{graphicx}  % got figures? uncomment this
\title{Non-perturbative QCD Effects and the Top Mass at the Tevatron}
\author{Daniel Wicke\from{ins:W}
\atque
Peter Z. Skands\from{ins:C}%
}
\instlist{\inst{ins:W} Bergische Universit{\"a}t Wuppertal \-
  Gau\ss{}str. 20, 42097 Wuppertal, Germany
\inst{ins:C}   CERN PH/TH  \- CH-1211 Geneva 23, Switzerland \\
  Theoretical Physics, Fermilab  \- MS106, Box 500, Batavia IL-60510, USA
}
\PACSes{
\PACSit{12.38.-t}{Quantum Chromodynamics}
\PACSit{13.85.Hd}{Inelastic scattering: many-particle final states}
\PACSit{13.87.Fh}{Fragmentation into hadrons}
\PACSit{14.65.Ha}{Top quarks}
}
\newlength{\ziffer}

\newcommand{\TeV}{\,\mbox{Te\kern-0.2exV}}
\newcommand{\GeV}{\,\mbox{Ge\kern-0.2exV}}
\newcommand{\mGeV}{\,\mathrm{Ge\kern-0.2exV}}
\newcommand{\MeV}{\,\mbox{Me\kern-0.2exV}}
\newcommand{\keV}{\,\mbox{ke\kern-0.2exV}}
\newcommand{\eV}{\,\mbox{e\kern-0.2exV}}

\newcommand{\bea}{\pagebreak[3]\begin{samepage}\begin{eqnarray}}
\newcommand{\eea}{\end{eqnarray}\end{samepage}\pagebreak[3]}
\newcommand{\beq}{\begin{equation}}
\newcommand{\eeq}{\end{equation}}

\newcommand{\abb}{Fig.~\ref}
\newcommand{\fig}{\abb}

 \newlength{\howlong}

\begin{document}

\maketitle

\begin{abstract}
The modelling of non-perturbative effects is an important part of
modern collider physics simulations. In hadron collisions there is
some indication that the modelling of the interactions of the beam
remnants, the underlying event, may require non-trivial colour
reconnection effects to be present.  
We recently introduced a universally applicable toy model of such
reconnections, based on hadronising strings. This model, which has one free
parameter, has been implemented in the Pythia event generator. We
then considered several parameter sets (`tunes'), 
constrained by fits to Tevatron minimum-bias data, and 
determined the sensitivity of a simplified top mass
analysis to these effects, in exclusive semi-leptonic top events
at the Tevatron. A first attempt at
isolating the genuine non-perturbative effects gave an estimate
of order $\pm0.5\GeV$ from non-perturbative uncertainties.
The results presented here are an update to the original study and include
recent bug fixes of Pythia that influenced the tunings investigated.
\end{abstract}
\section{Introduction}
The top quark mass is the only free parameter specific to the top
quark sector of the Standard Model of elementary particle physics (SM). Direct
measurements from the Tevatron~\cite{Group:2008nq} combined with indirect
determinations from electroweak precision measurements can therefore be 
used to test the consistency of the SM and to predict the Higgs boson mass
within this theoretical framework~\cite{Alcaraz:2007ri}. 

The question of whether the direct and the indirect results concern 
the same mass parameter is an important issue in this context.
At the very least, the same theoretical definition must be used
throughout, to ensure that consistency checks and Higgs mass
predictions are valid.  
At present, it is customary to assume that the quoted values for
direct measurements correspond to the pole 
mass. In practice, all direct measurements of the top quark mass are
calibrated back to a value corresponding to 
the input top quark mass in one of the Monte
Carlo programs~\cite{Sjostrand:2006za,Corcella:2000bw}. Uncertainties
on higher order corrections and soft QCD effects affect this
calibration, through the modelling of parton showers, underlying event,
colour reconnections, and hadronisation. One can therefore be
concerned whether these calibrations alter the meaning of the
mass determined in Tevatron mass measurements.  

In an earlier work \cite{Skands:2007zg}, we addressed the influence of
the modelling used in simulating 
top quark pair events in Tevatron proton anti-proton collisions with special
emphasis on underlying-event and colour-reconnection effects. The
influence of various models on a toy mass measurement was used to derive
calibration uncertainties for the top quark mass measurement from
non-perturbative QCD effects. 

We here update the results presented in \cite{Skands:2007zg}. Most
importantly, we use a more recent version of the Pythia generator
(6.416, as compared to 6.408), which takes into account 
important bug fixes affecting the tuning of the $p_\perp$-ordered
parton shower (see \cite{pythia-update-notes}). We also include a
comparison to recently published CDF data on the average $p_\perp$ as
a function of observed multiplicity in minimum-bias events \cite{cdf}, a
distribution which appears to be highly sensitive to colour correlations. 

\section{Modelling in Hadron Collisions}\label{modelling}
The simulation of proton anti-proton collision at the Tevatron is separated
into many steps, only some of which are computable from first principles.
The parton content of the colliding protons and anti-protons is described by
parton density functions fitted to experimental data. At a ``hard
scale'' characteristic of the process in question, these functions are
convoluted with matrix elements to describe the fundamental
short-distance process. Initial-state 
parton showers then evolve the incoming partons selected for the hard
interaction ``backwards'' \cite{Sjostrand:1985xi} from the hard scale
down to an infrared cutoff, and likewise final-state showers evolve
the outgoing partons down to a hadronisation cutoff, at
which point a hadronisation model takes over. 
In parallel to the hard process, additional interactions (gluon
exchanges) may occur
between the (anti-)proton remnants. These give rise to an ``underlying
event'' (UE), whose detailed structure cannot at present be derived
from first principles. Instead, generators like Pythia
\cite{Sjostrand:2006za} and Jimmy \cite{Butterworth:1996zw}
rely on models of multiple parton-parton interactions (MPI)
\cite{Sjostrand:1987su,Butterworth:1996zw,Sjostrand:2004ef} to describe
this aspect.  

\begin{figure}[t]
  \centering
\includegraphics[width=0.5\linewidth,clip,trim=0mm 1mm 0mm 0mm]{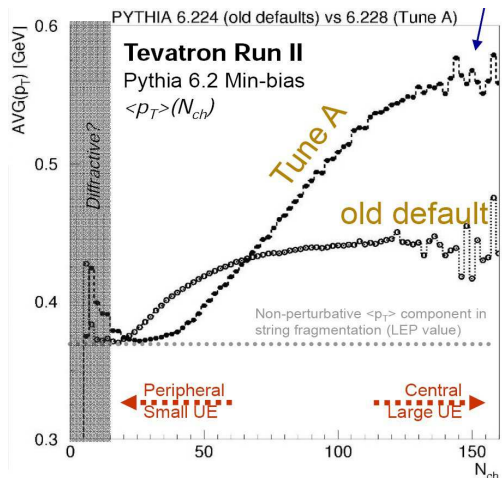}  
  \caption{Average transverse momentum in minimum bias events as function of
    the charged multiplicity. One can clearly see the change between the old
    default and Tune A.
}
  \label{fig:TuneA}
\end{figure}

The Monte Carlo generator Pythia 6.4~\cite{Sjostrand:2006za} actually 
provides two different, but related, 
models to describe the underlying event. The
older model \cite{Sjostrand:1987su} treats the 
underlying event only after initial-state showering of the hard process is
complete. Additional back-to-back parton pairs 
(either $gg$ or $q\bar{q}$, with the relative
composition user-specifiable) are then allowed to be created 
by subsequent MPI processes among the remnants. These extra parton 
pairs are not showered, but are fed  directly into the hadronisation,
together with the 
partons from the hard process. The colour correlations between the
hard-process partons and the MPI partons are
user-specifiable. Empirically, they are found to have to be rather
strong, see below.
The new model \cite{Sjostrand:2004ef} instead interleaves the
additional MPI processes, and showers off them, 
with the initial-state parton shower evolution off the hard process. 
It thereby implements a successive fine-graining of all perturbative
activity. Various options for colour connections and colour
reconnections between and inside the MPI ``chains'' exist, see
below. The new model also incorporates a more sophisticated treatment
of beam remnants \cite{Sjostrand:2004pf}, including baryon junctions
\cite{Sjostrand:2002ip}.

The models are governed by several 
parameters influencing, e.g., the probability of
additional interactions or their momentum distributions. These parameters are
not known a priori and thus need to be tuned to describe the data.
We have here chosen to focus on two specific distributions, the charged
multiplicity and the average transverse momentum of the particles in events
selected with a minimum-bias trigger.
It should be noted that, even with tuning, one should not expect all
models to be able to describe the data, owing to shortcomings in their
physics descriptions, as will be commented on below. 

Several older tunes of Pythia parameters 
obtained from CDF fits to minimum bias data are
available, e.g., Tune A,  Tune DW,  Tune BW, etc.\cite{tunea}.
These all significantly modify the original default parameters for the
underlying event, c.f.\ \fig{fig:TuneA}, changes which are 
motivated directly by improving the description of the data.
One striking common feature of these tunes is that the parameters describing
the probability of non-trivial colour connections between 
the additional-parton interactions and the hard scattering, {\tt
PARP(85)} and {\tt PARP(86)}, are significantly enhanced.
We here interpret this as a sign of actual colour
reconnections happening in the underlying event, and investigate
the consequences of this hypothesis.

\section{Colour Reconnection Models}\label{models}
In hadron collisions, the underlying event
produces an additional amount of displaced colour charges, translating
to a larger density of hadronising strings between the beam
remnants. It is not known to what extent the collective hadronisation
of such a system differs from a sum of independent string pieces. 
Measurements at LEP
\cite{Abbiendi:1998jb,Abbiendi:2005es,Schael:2006ns,Abdallah:2006ve} 
would not have been sensitive to this effect, and hence it is quite
possible that colour reconnection (CR) effects in hadron collisions
may be substantially stronger than the LEP constraints 
would appear to allow, if taken at face value.

However, most of the CR models investigated at LEP focused 
exclusively on $WW$ physics, and so were not immediately applicable to 
hadron collisions. Colour reconnection effects in $t\bar{t}$ events
were first considered in \cite{Khoze:1994fu}, but also there 
only in the context of $e^+e^-$ collisions. 
We therefore recently introduced a toy model of 
colour reconnection models for more general situations, based on an
annealing-like minimisation of a measure of the potential energy of
the confinement field. This so-called colour annealing
model \cite{Sandhoff:2005jh} has been implemented in the Pythia
generator since version 6.402. Alternative models, e.g.\ the ones by
Rathsman~\cite{Rathsman:1998tp} and by Webber~\cite{Webber:1997iw},
would also be interesting to explore, but we have so far not done
this. 
\begin{figure}[t]
  \centering
  \includegraphics[width=0.49\textwidth,clip,trim=0mm 5mm 0mm 15mm]{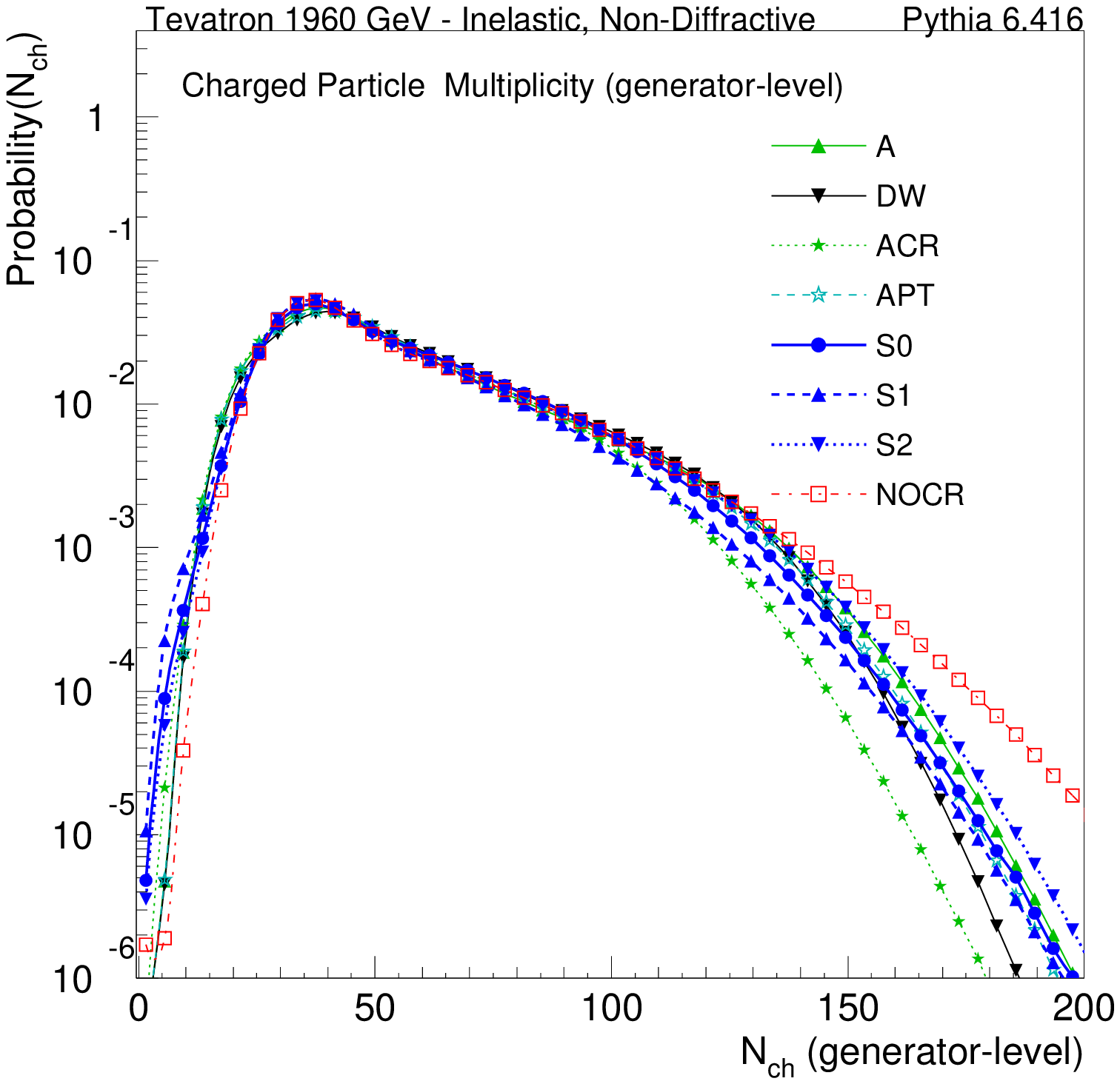}
\includegraphics[width=0.49\textwidth,clip,trim=0mm 5mm 0mm 15mm]{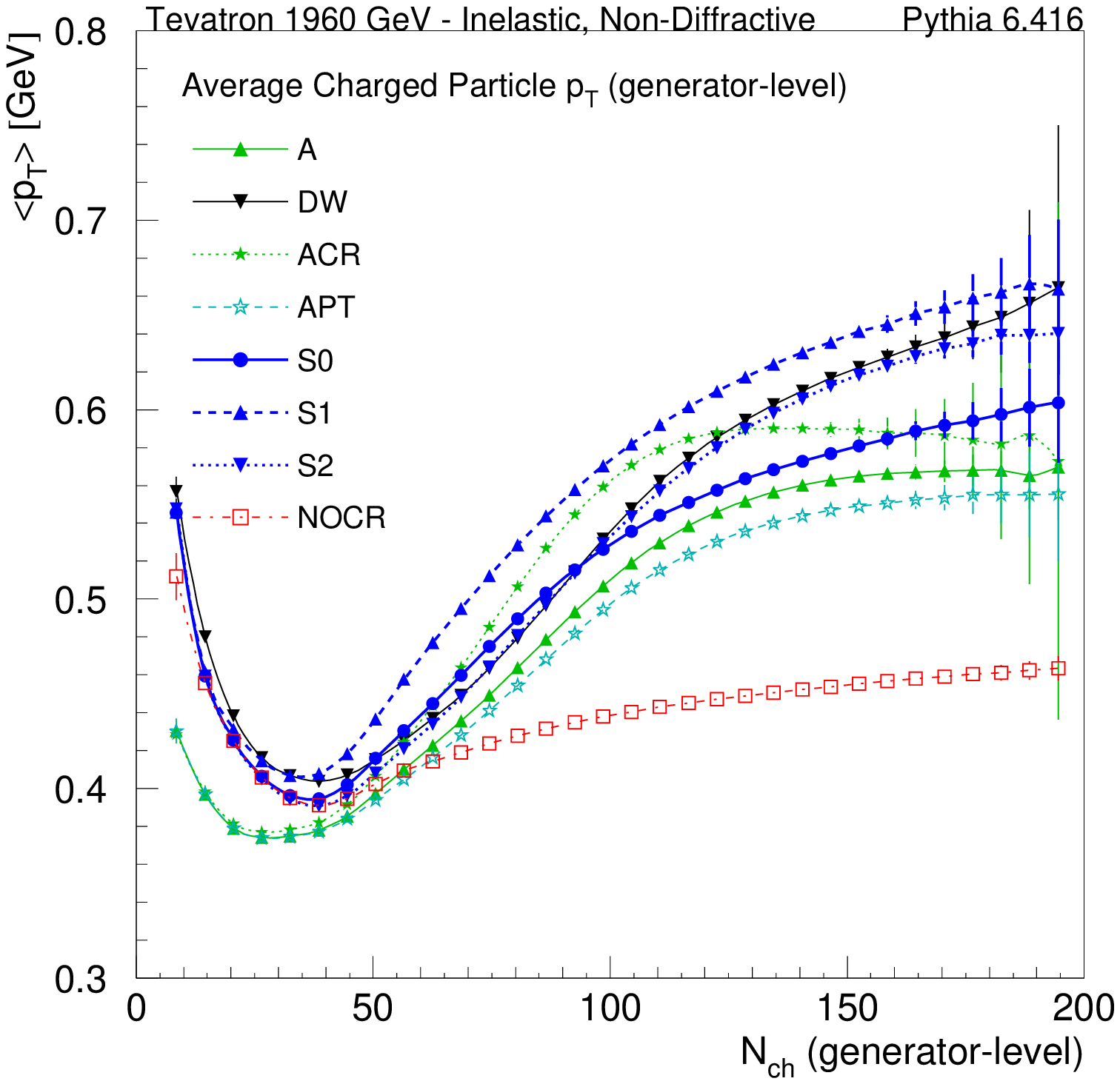}
  \caption{Generator level comparison of various models available in Pythia 6.416. Charged
    multiplicity distribution (right) and mean transverse momentum as function
    of the charged multiplicity. }
  \label{fig:tuneobservables}
%\end{figure}
%\begin{figure}
  \centering
  \includegraphics[width=0.49\textwidth,clip,trim=0mm 5mm 0mm 15mm]{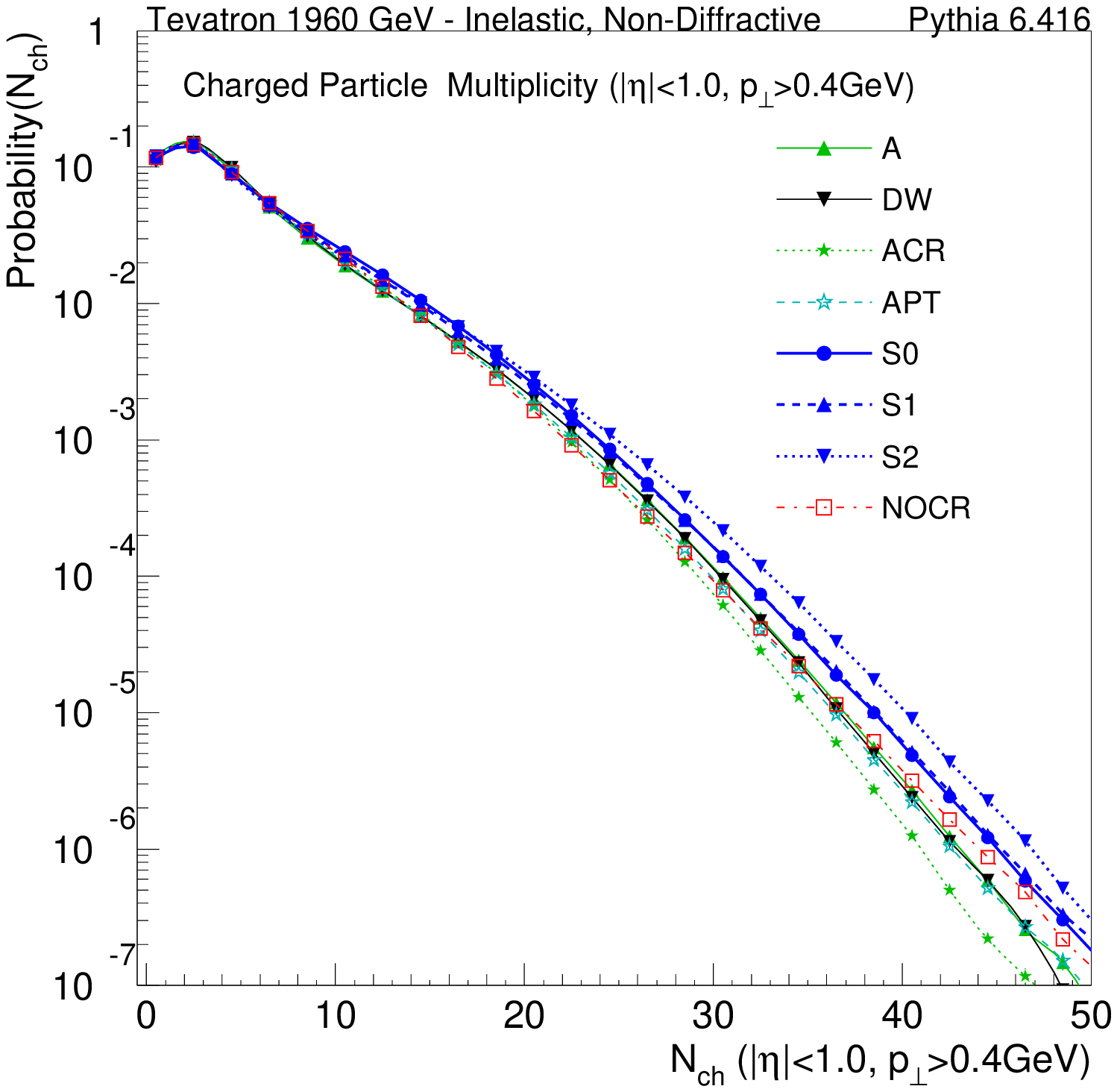}
\includegraphics[width=0.49\textwidth,clip,trim=0mm 5mm 0mm 15mm]{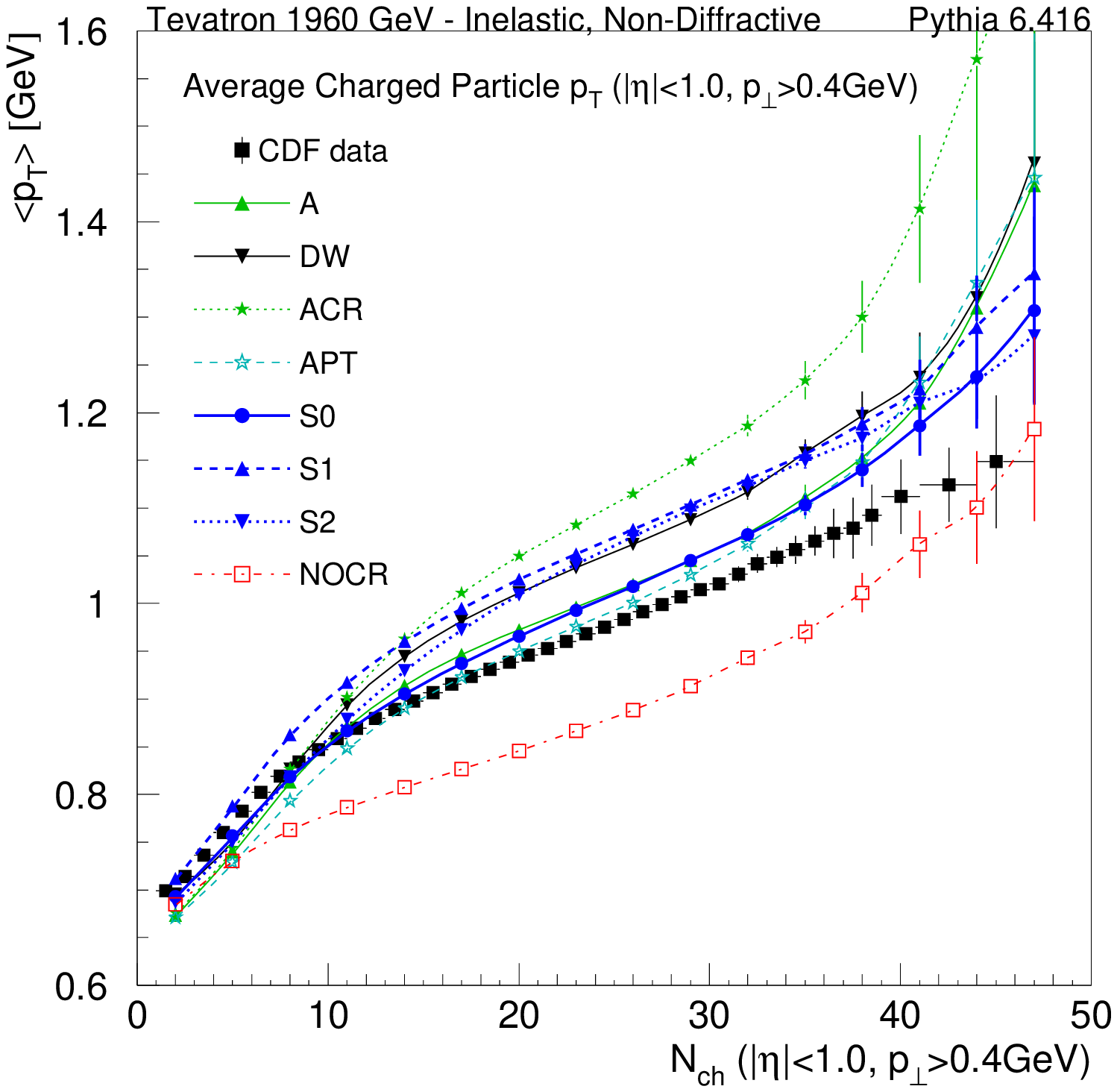}
  \caption{Comparison of various models available in Pythia 6.416 including
    cuts used by CDF ($p_\perp>0.4\GeV$, $\left|\eta\right|<1.0$)~\cite{cdf}. 
    Charged
    multiplicity distribution (right) and mean transverse momentum as function
    of the charged multiplicity. The data shown in the right
    plot became available during the conference and so 
    were not yet used directly to tune the annealing models.
  }
  \label{fig:tuneobservablescdf}
\end{figure}
We emphasise that both in the old models (Tune A and cousins) 
as well as in the new models (Tune S0 and cousins) the colour
correlations of the underlying event affects the string topology of
the hard interaction as well. In the old model, gluons from the
underlying event are sequentially ``attached'' to string pieces
defined by the hard interaction, and hence cause wrinkles and kinks on
the existing topology. In the new, annealing, models, more radical changes are
possible, with the colour flow of the hard interaction not necessarily
preserved at all.

Briefly described, the annealing models work in the following way
\cite{Sandhoff:2005jh}.  
At hadronisation time, each string piece is a assigned a fixed probability
to participate in the annealing,
$P_\mathrm{reconnect}=1-(1-\chi)^n$. Here, $\chi$  
is parametrising the strength of the reconnection effect and $n$ is the number
of parton interactions, i.e. roughly 
counts the number of possible reconnections.
For the participating string pieces, new connections are chosen to
minimise the string length and thus to minimise the potential energy
in the string. Note that these new 
connections are not explicitly prevented from
being the same as the original ones, if that is what minimises the
string length. 
Pythia provides three model variations that differ in the
suppression of gluon-only strings.

The colour annealing models yield important changes in the description of
the underlying event. Thus both the CR and UE models were tuned anew. 
Due to difficulties in obtaining the data itself, the
tuning was originally done with respect to two of the existing model parameter
sets: Tune A and Tune DW. Two observables,
the charged multiplicity distribution and the average transverse momentum as
function of the charged multiplicity, $\langle p_T \rangle(N_\mathrm{ch})$,
were used for the tuning. 
In \fig{fig:tuneobservables} the  various model tunes are compared for the
two observables used in tuning. Most of the models give very similar
results, only the {\tt 
  NOCR} model differs strongly in $\langle p_T \rangle(N_\mathrm{ch})$. This
model explicitly uses no colour reconnection at all. 
This underlines the fact that CR appears to be a necessary ingredient 
to achieve consistency with the data. 
A measurement of $\langle p_T \rangle(N_\mathrm{ch})$ by CDF~\cite{cdf} became
available during the conference. It uses additional cuts on the track
momenta and 
pseudo-rapidity range, reflecting the CDF detector acceptance. 
Results with these cuts and a comparison to data is
shown in \fig{fig:tuneobservablescdf}.

Tunes for the described colour reconnection models first appeared in Pythia
v6.408 and were revised in v6.414 after a  bug affecting the $p_T$
ordered shower was fixed. Since v6.413, hard-coded presets for all
tunes considered 
here, and others, can be accessed via Pythia's {\tt MSTP(5)} variable,
see \cite{pythia-update-notes}.  Some additional discriminating 
distributions and extrapolations to the LHC can be found in
\cite{Skands:2007zz}. 

\section{Toy Top Mass Measurements}\label{toymass}

As discussed in the introduction, the underlying event and colour
reconnection effects may influence the results 
obtained in measurements of the hard process. At LEP, the $W$ mass was
especially sensitive to these effects. This brings us back to the question of 
the influence of the various models on measurements of the top quark
mass at the Tevatron. 

Current measurements of the top quark~\cite{Group:2008nq} consist of
three main ingredients: 
First, a mass estimator based on the reconstructed physics objects,
i.e., jets, lepton and missing transverse energy. Such an estimator
uses a jet-parton 
assignment done by either 
choosing or weighting the various possibilities. Second, current
measurements include 
an overall jet energy scale (JES) correction factor, which reduces the
dominating 
systematic uncertainty by using the well known $W$
mass as an additional constraint. 
And finally all methods are calibrated to simulation by correcting any
offset between the reconstructed top mass and the nominal value of the
simulation. It is especially in this last step that
the different models may affect the outcome
of the procedure. 

To concentrate on physics effects and to avoid dealing with detector
simulation a simplified toy mass measurement for lepton plus 
jets events is implemented on generator level. 
For this toy mass measurements only semi-leptonic top pair events are
investigated. Jets are reconstructed using a cone
algorithm~\cite{ktjet,plano} with $\Delta 
R=0.5$, $p_T>15\GeV$. The events are required to have exactly four such
jets. A simplified jet-parton assignment is done by matching the reconstructed
jets to the Monte Carlo truth by $\Delta R$. Only events with a unique
assignment are kept. 
The top mass is computed in each event from the three jets assigned to the
hadronically decaying top quark.

To obtain a mass estimator for a full dataset the peak of the the distribution
of reconstructed top mass values is fitted with a Gaussian. A fit range of
$\pm15\GeV$ is used and the fit is iterated to assure that the final fit range
is symmetric around the fitted mass, $m_\mathrm{top}^\mathrm{fit}$.
As the jets aren't corrected for out-of-cone effects this mass estimator is
expected to give results that are lower than the nominal value in the
simulation. In analogy to the JES correction factor this can be corrected for
by using the $W$ mass information.  An event-by-event $W$ mass is
reconstructed again by 
fitting a Gaussian to the distribution of mass values reconstructed from the
two jets assigned to the hadronic $W$ decay. A scaled top mass estimator is
then constructed as 
$m_\mathrm{top}^\mathrm{scaled}=s_\mathrm{JES}\,m_\mathrm{top}^\mathrm{fit}$
where the scale factor is $s_\mathrm{JES}=80.4\GeV/m_W$, 
with $m_W$ being the $W$
mass obtains from the fit.
Thus the simplified top mass measurements provide two results
$m_\mathrm{top}^\mathrm{fit}$ and $m_\mathrm{top}^\mathrm{scaled}$, one
before JES correction and one after.

\section{Calibration Uncertainties for the Top Mass}\label{calibration}
\begin{figure}[b]
  \centering
\unitlength=0.73mm
\begin{picture}(80,73)
\put(17,40){\includegraphics[width=30\unitlength]{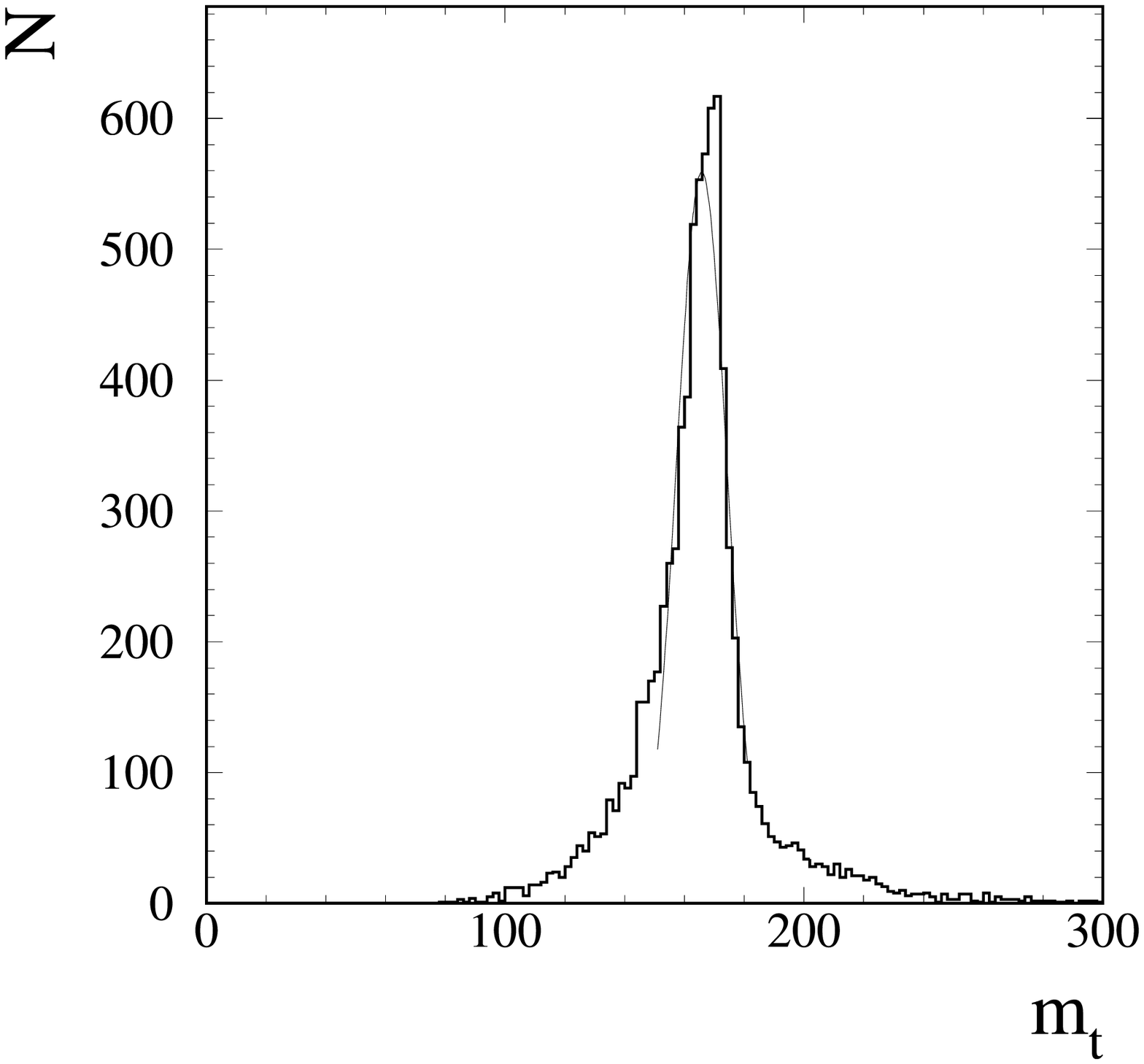}}
\put(0,0){\includegraphics[width=80\unitlength,clip,trim=0mm 0mm 0mm 3.8mm]{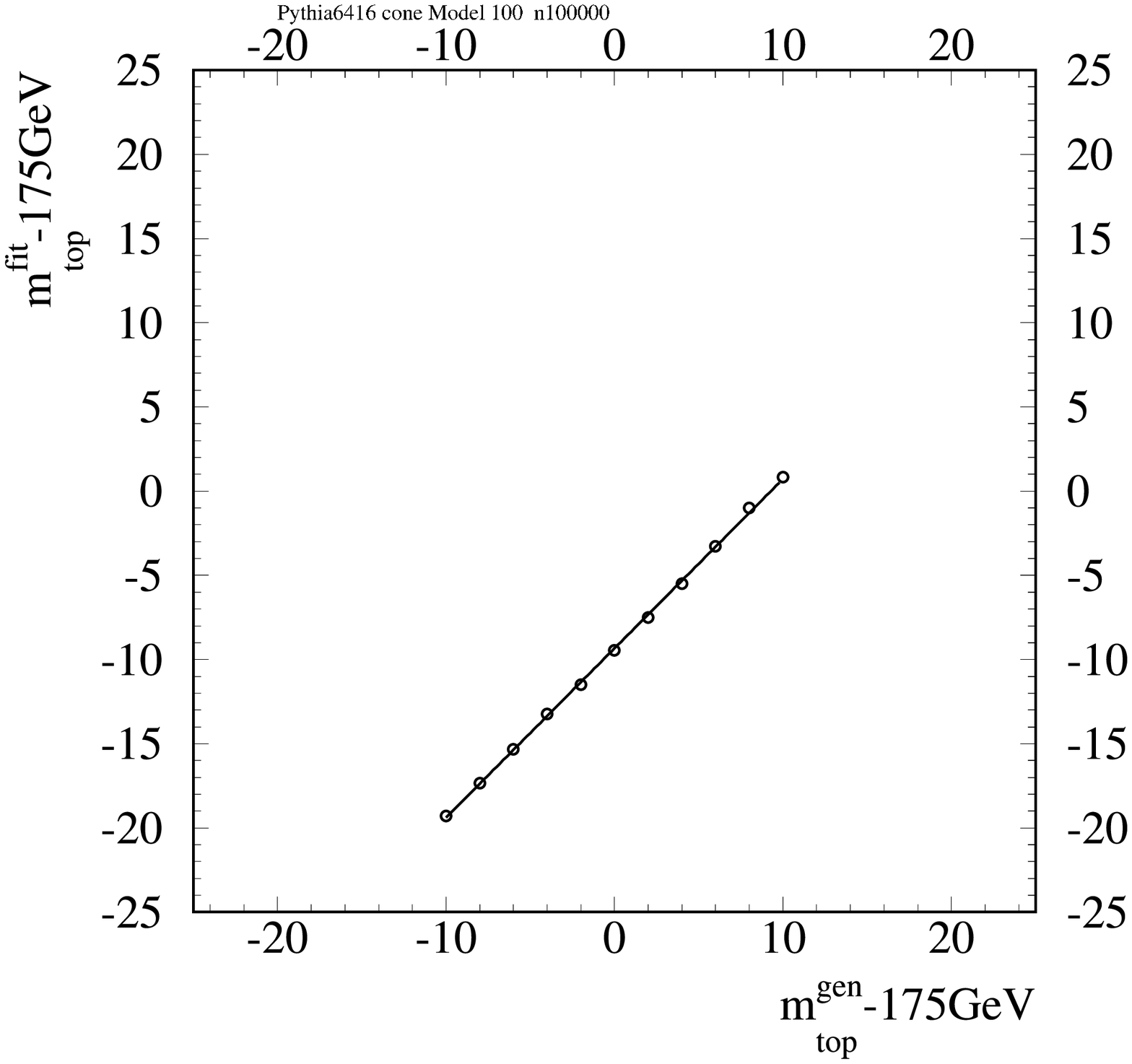}}
\end{picture}
\hfill
\raisebox{40\unitlength}{$\begin{array}{c}{\mathrm{\small rescale}}\\{\longrightarrow}\end{array}$}
\hfill
\includegraphics[width=80\unitlength,clip,trim=0mm 0mm 0mm 3.8mm]{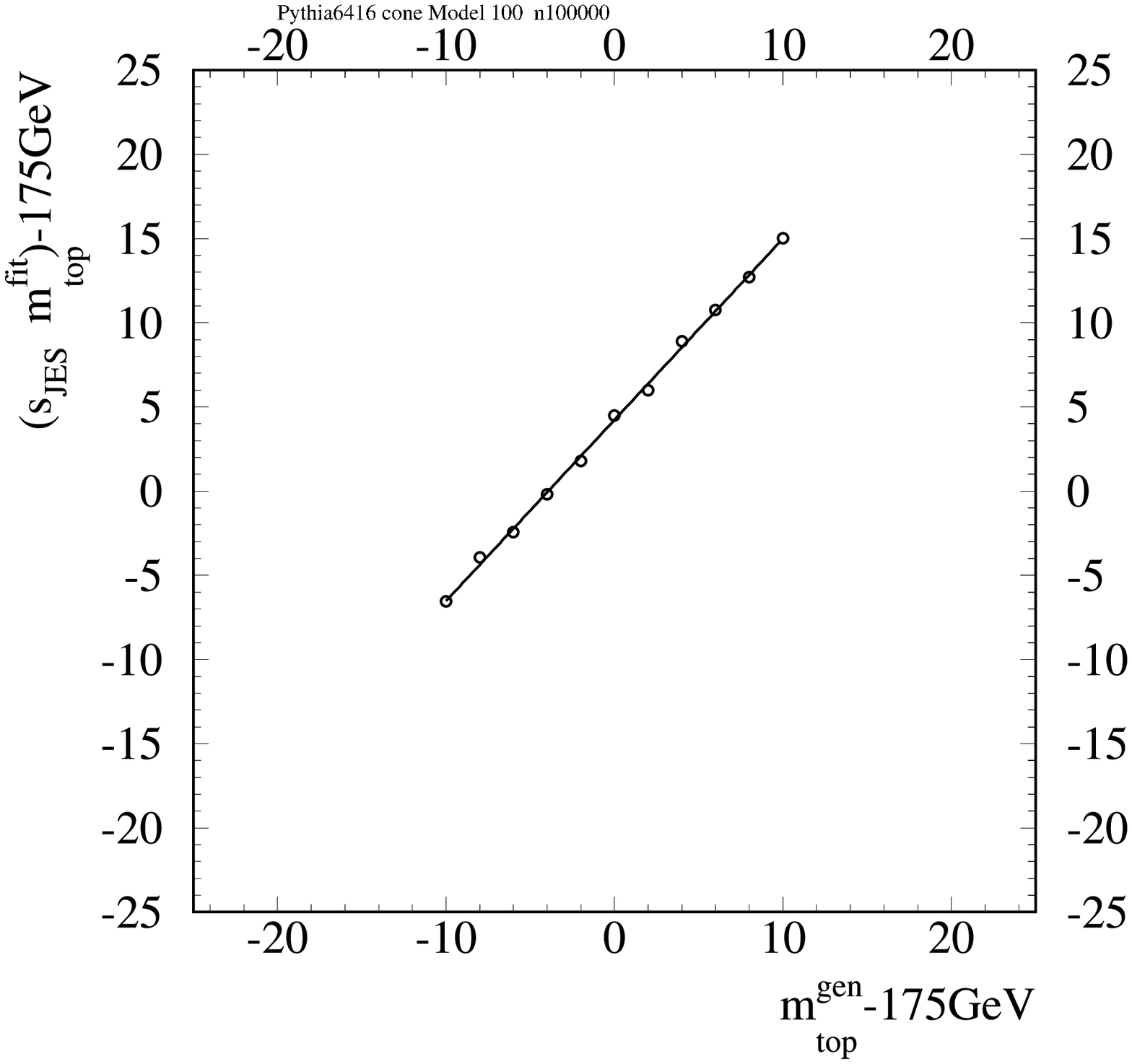}\hfill\null
  \caption{Calibration curve obtained with Tune A before (left) and after
    (right) JES rescaling. The inset shows the Gaussian fit to the
    distribution of reconstructed top masses for one specific nominal top
    mass, $m_\mathrm{top}^\mathrm{gen}=175\GeV$. 
  }
  \label{fig:calibration}
\end{figure}

\begin{figure}[t] 
  \centering
\unitlength=0.85mm
\begin{picture}(0,63)
\put(10.2,63){$\displaystyle\overbrace{\hspace*{8\unitlength}}^{\Delta m_\mathrm{top}^\mathrm{fit}}$}
\put(67,63){$\displaystyle\overbrace{\hspace*{10\unitlength}}^{\Delta m_\mathrm{top}^\mathrm{scaled}}$}
\end{picture}
\includegraphics[width=84\unitlength,clip,trim=0mm 5mm 0mm 7mm]{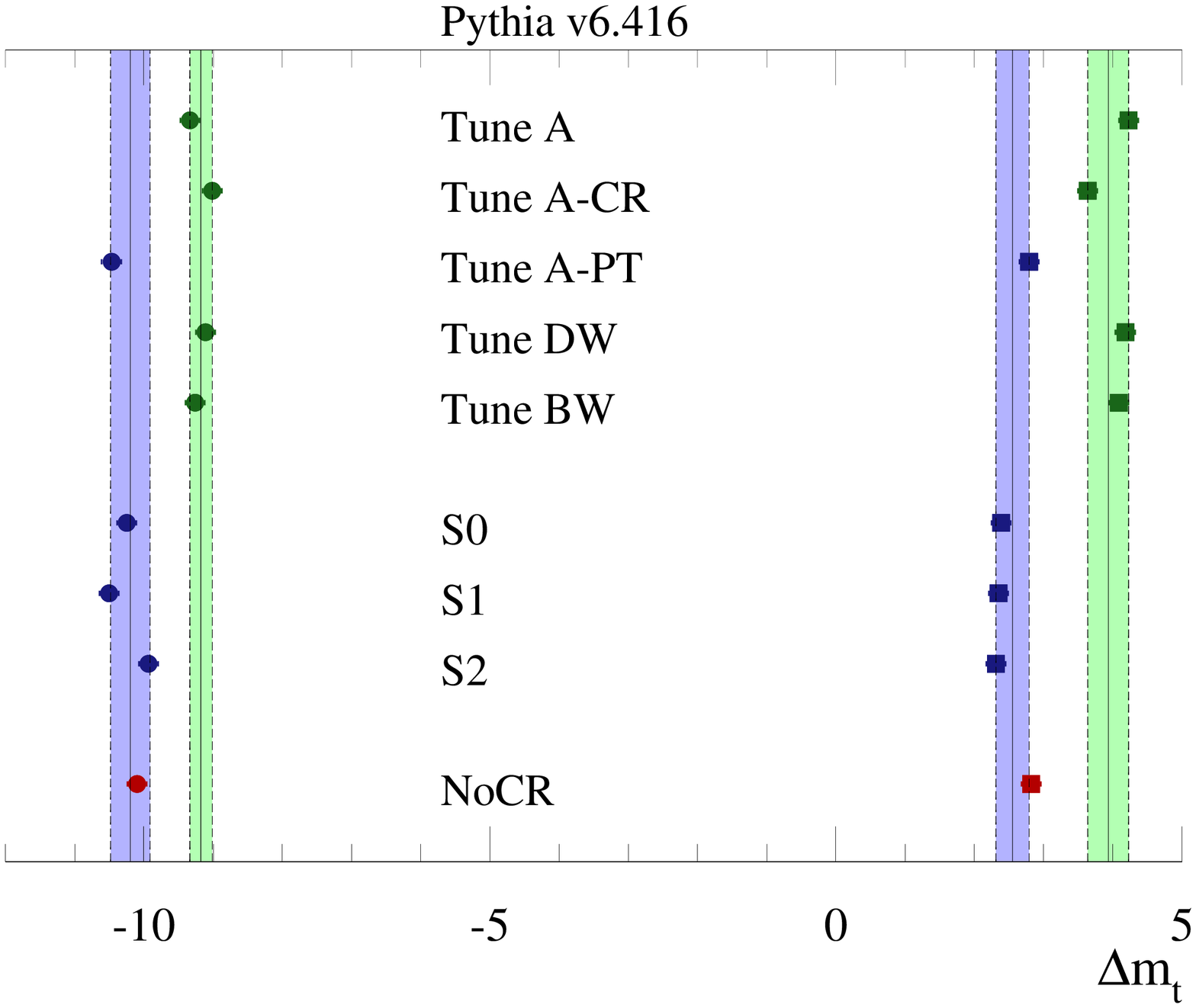}
  \caption{Comparison of calibration offsets obtained for each model.
The column on the left (dots) show the results obtained before $\mathrm{JES}$
rescaling, the right column (squares) after rescaling. The statistical
precision due to the finite number of generated events is at the level of
$\pm 0.15\GeV$.}
    \label{fig:offsets}
\end{figure}

Calibration curves for both top mass estimates were created by scanning the
nominal top mass from $165$ to $185\GeV$ and computing the described mass
estimator for each of the nominal values. Example calibration curves for both
mass estimators are shown in \fig{fig:calibration}. The curves show an
excellent linearity. As expected the offset for the un-scaled estimator,
$m_\mathrm{top}^\mathrm{fit}$, is negative and scaling with the $W$ mass
brings the scaled result, $m_\mathrm{top}^\mathrm{scaled}$, closer to the nominal result.
The procedure has been
repeated for various (tuned) models to compare the offsets of the calibration curves.
The offsets are obtains from a straight line fit and evaluated at
$m_t=175\GeV$. The results for the various models are summarised in
\fig{fig:offsets} for both mass estimators. 
The models exhibit a spread of $\pm 0.8\GeV$ and $\pm 1.0\GeV$ for the 
$m_\mathrm{top}^\mathrm{fit}$ and $m_\mathrm{top}^\mathrm{scaled}$ estimators,
respectively. 

In real mass measurements the calibration offsets are used to correct the
mass estimator to the nominal value. The corrections derived from the
simulation are then applied to the real data. But the choice of the
model used to perform the calibration is ambiguous and hence 
the spread between the
models must be considered as a calibration uncertainty.
The source of the spread can be separated into two sources by noting that the
models used fall in two classes: Those that utilise the 'old'
virtuality-ordered 
parton shower and those that utilise the 'new' $p_T$-ordered one. The largest
component of the difference is \em between \em these two classes, indicating a
perturbative nature of most of it. Within each class differences of
less than $\pm0.5\GeV$ on the top mass remain, which are assigned to the
non-perturbative differences between the various models. In \fig{fig:offsets}
the classes are grouped by coloured bands.

It should be noted that different mass estimators may have a different sensitivity to the
model differences and thus may exhibit a different uncertainty. The results of
this toy mass analysis are therefore only a first hint to the actual size of
the effects, which should be studied for each real mass measurement separately.

\section{Summary}\label{summary}
Top mass measurements are now reaching total uncertainties below $1.5 \GeV$. At
this precision non-perturbative effects may become important. A set of new,
universally applicable models to study colour reconnection effects in hadronic
final states was presented.  The models apply an
annealing-like algorithm that minimises the potential energy within string
hadronisation models. The models were tuned simultaneously with the
underlying-event  description of Pythia to distributions sensitive to
non-perturbative effects in minimum-bias samples.
The influence of changing underlying event model, the colour reconnection and
parton showers on measurements of the top mass was investigated in a toy mass
analysis, resulting in variations of about $\pm1.0\GeV$ on
the reconstructed top mass. Of this total uncertainty we tentatively
attribute about
$0.7\GeV$ to perturbative effects and of less than $0.5\GeV$ to non-perturbative
sources. These results were obtained with Pythia v6.416 with tunes
updated after 
fixing a bug in the $p_T$ ordered shower. While the model differences are
slightly reduced with the new version of Pythia, the qualitative conclusions
of~\cite{Skands:2007zg}, 
derived with an older version of the generator and tunes, remain unchanged.

\section*{Acknowledgements}
PS is supported by the Fermi Research Alliance, under contract DE-AC02-07CH11359 
with the U. S. Dept. of Energy, and by the European Union Marie Curie Research 
Training Network MCnet under contract MRTN-CT-2006-035606.
\bibliographystyle{varenna-DW}
\begin{flushleft}
\bibliography{CR,letter}
\end{flushleft}
\end{document}